\documentclass[preprint]{elsarticle}
\usepackage[detect-all]{siunitx}
\usepackage[utf8]{inputenc}
\usepackage{subcaption}
\usepackage{enumitem}
\usepackage{isotope}
\setlist{nolistsep,nosep}
\usepackage{amsmath}
\usepackage{hyperref}

\DeclareSIUnit\ele{e^{\text{-}}}

\journal{Nucl. Instrum. Methods Phys. Res. A}
\begin{document}

\begin{frontmatter}

\title{Charge collection properties of TowerJazz \SI{180}{\nm}~CMOS Pixel Sensors in dependence of pixel geometries and bias parameters, studied using a dedicated test-vehicle: the Investigator chip}

\author[1]{G. Aglieri Rinella}
\author[2]{G. Chaosong}
\author[1]{A. di Mauro}
\author[3]{J. Eum}
\author[1]{H. Hillemanns}
\author[1]{A. Junique}
\author[1]{M. Keil}
\author[4]{D. Kim}
\author[3]{H. Kim}
\author[1]{T. Kugathasan}
\author[3]{S. Lee}
\author[1]{M. Mager}
\author[5]{V. Manzari}
\author[1]{C.~A. Marin Tobon}
\author[1]{P. Martinengo}
\author[6]{H. Mugnier}
\author[1]{L. Musa}
\author[1]{F. Reidt}
\author[6]{J. Rousset}
\author[1]{K. Sielewicz}
\author[1]{W. Snoeys}
\author[1]{M. Šuljić \corref{cor1}}
\author[1]{J.~W. van Hoorne}
\author[1]{Q.~W. Malik}
\author[2]{P. Yang}
\author[3]{I.-K. Yoo}

\address[1]{European Organization for Nuclear Research (CERN), Geneva, Switzerland}
\address[2]{Central China Normal University, Wuhan, China}
\address[3]{Department of Physics, Pusan National University, Pusan, Republic of Korea}
\address[4]{Department of Semiconductor Science, Dongguk University, Seoul, Republic of Korea}
\address[5]{INFN, Sezione di Bari, Bari, Italy}
\address[6]{MIND, Archamps, France}

\cortext[cor1]{Corresponding author}

\begin{abstract}
This paper contains a compilation of parameters influencing the charge collection process extracted from a comprehensive study of partially depleted Monolithic Active Pixel Sensors with small (\SI{<25}{\um^2}) collection electrodes fabricated in the TowerJazz \SI{180}{\nm}~CMOS process. These results gave guidance for the optimisation of the diode implemented in ALPIDE, the chip used in the second generation Inner Tracking System of ALICE, and serve as reference for future simulation studies of similar devices. The studied parameters include: reverse substrate bias, epitaxial layer thickness, charge collection electrode size and the spacing of the electrode to surrounding in-pixel electronics. The results from pixels of \SI{28}{\um} pitch confirm that even in partially depleted circuits, charge collection can be fast (\SI{<10}{\ns}), and quantify the influence of the parameters onto the signal sharing and amplitudes, highlighting the importance of a correct spacing between wells and of the impact of the reverse substrate bias.
\end{abstract}


\end{frontmatter}


\section{Introduction}
\label{sec:introduction}

\begin{figure}[!ht]
    \centering
    \includegraphics[width=1\textwidth]{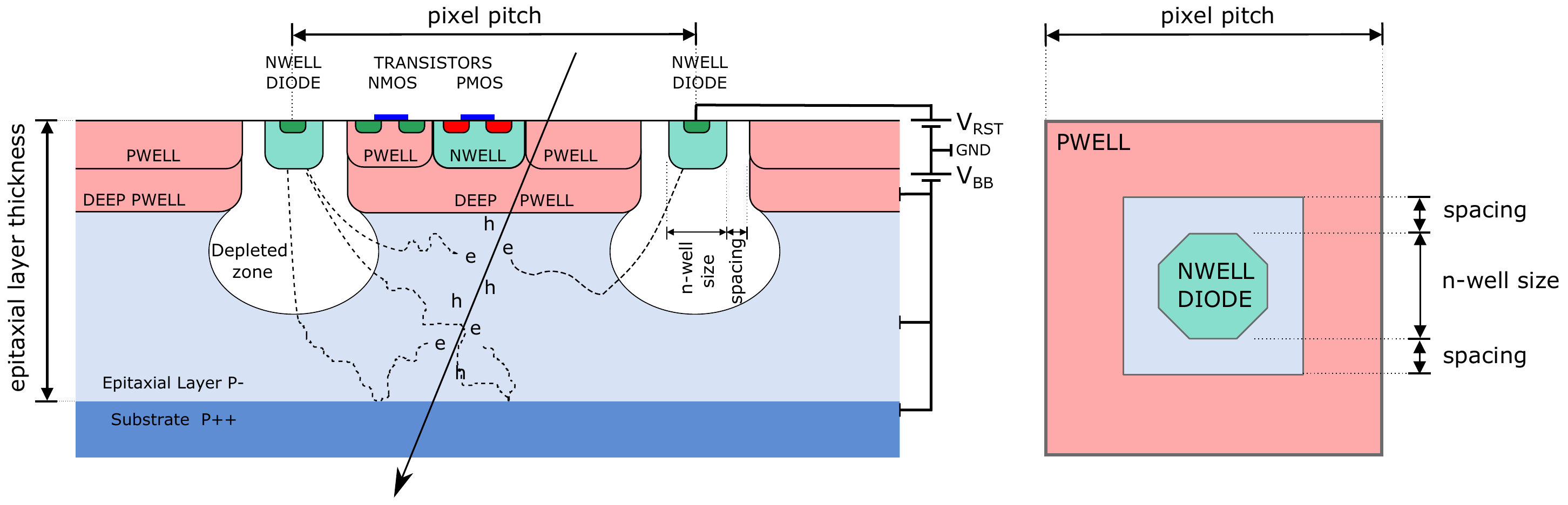}
    \caption{Schematic cross-section and top-view (not to scale) of the well structure used in the ALPIDE and Investigator chips, showing the key design parameters that influence the charge collection.}
    \label{fig:pixel}
\end{figure}

During the R\&D for the new Inner Tracking System~(ITS)~\cite{TDR} of the ALICE experiment at CERN LHC, a novel Monolithic Active Pixel Sensor (MAPS), named ``ALPIDE'', was developed by the collaboration~\cite{ALPIDE-proceedings-1, ALPIDE-proceedings-2,ALPIDE-proceedings-3}. The chip is produced in \SI{180}{\nm}~CMOS technology of TowerJazz\footnote{\url{http://www.towerjazz.com/}} and features a quadruple well structure, which allows the use of PMOS together with NMOS devices within the pixel matrix. The schematic structure of a pixel cell in this technology is illustrated in Fig.~\ref{fig:pixel}. 

For the ALPIDE, the ALICE development has focused on partially depleted MAPS with small collection electrodes, implemented in the standard TowerJazz process.
A process modification to fully deplete the sensitive layer was also introduced as a side development~\cite{modified,modified2}, followed by additional modifications to further accelerate the charge collection in the context of detector developments for the Compact Linear Collider (CLIC)~\cite{clictd,clictdyellow,clictdTWEPP} and to enhance radiation tolerance for applications in  ATLAS~\cite{atlas,malta,malta2,minimalta}.

For partially depleted devices, the generated electrons are transported by both diffusion and drift before being collected by the strong drift field in the depleted zone around the small collection electrode.
A relatively large voltage signal of $\Delta V\approx\SI{100}{\mV}$ is generated on the collection electrode due to its small capacitance $C$, typically of the order of few~\si{\fF} \cite{jacobus_thesis} ($\Delta V=Q/C$, with $Q$ being the collected charge). From this argument, it can already be seen that the performance will depend crucially on the spatial extension of the depleted region (or more generally the strength of the electric field) as it influences both the diode capacitance and the amount of diffusion and hence the charge spread. Finally, it will directly determine the duration of the collection process.

As a precise sensor technology CAD simulation model requires detailed knowledge of doping profiles for the entire sensor (wells, epitaxial layer, substrate), a series of different detection geometries were prototyped, both as benchmarks for simulations and for an heuristic approach to the optimisation of pixel geometries.

\section{The Investigator chip}
\label{sec:investigator}

The Investigator chip was developed to study the MAPS design parameter space in the context of the ALICE ITS upgrade.
It is produced using the TowerJazz \SI{180}{\nm}~CMOS imaging process on wafers with a high-resistivity (\SI{>1}{\kilo\ohm\cm}) epitaxial layer of three thicknesses: 18, 25 and~\SI{30}{\um}. 
A reverse substrate bias voltage ($V_{BB}$, cf.~Fig.~\ref{fig:pixel}) can be applied to the sensor\footnote{Breakdown is observed between \num{-8}~and~\SI{-10}{\V}. To ensure sufficient operational margin, the chips were operated down to $V_{BB}=\SI{-6}{\V}$.}.

The performance of a pixel matrix is a combination of different parameters, amongst which the most prominent are conversion gain (capacitance), charge collection time, charge spread, and total charge collection efficiency. To access these parameters quantitatively, it is necessary to measure the induced signal in a direct and time-resolved way, simultaneously on a number of collection diodes. 
The Investigator chip contains more than hundred matrices with pixel pitches ranging from 20 to \SI{50}{\um}, different collection diode geometries, reset mechanisms, and input-transistor configurations~\cite{Gao}.
Each of the matrices contains \num{10x10}~pixels, of which the central \num{8x8} are read out in parallel, in an analogue fashion. 

Due to its versatility, the Investigator also found applications as test vehicle in several other R\&D contexts; CLIC~\cite{clictd,clictdyellow,clictdTWEPP}, ATLAS~\cite{atlas}, and the study of modifications of the CMOS process\footnote{The chips studied in this paper use the standard process without custom modifications.} to increase its timing performance and radiation hardness~\cite{modified,modified2}. 

\subsection{Pixel geometries}
\label{sec:geometries}

Figure~\ref{fig:pixel} shows a schematic cross section of a pixel and indicates the key geometric parameters that influence the charge collection: pixel pitch, epitaxial layer height, diode n-well size and spacing between diode n-well and surrounding p-well. The Investigator chip implements several combinations of these parameters. While they have to obey some boundary conditions in a fully integrated chip like ALPIDE, mainly imposed by the area needed to implement the in-pixel circuitry (e.g.~discriminator, masking, in-pixel latches, readout network), the Investigator chip also contains pixel variants with larger diode n-wells and spacings, which is possible due to its minimal in-pixel circuitry.

Table~\ref{tab:geomtries} summarises the \num{13}~different geometries that where studied for this paper. The pixel pitch is fixed\footnote{ALPIDE eventually features a \SI{29.24x26.88}{\um} pitch due to global chip integration requirements~\cite{ALPIDE-proceedings-1,ALPIDE-proceedings-2,ALPIDE-proceedings-3}} to \SI{28}{\um} and two values of epitaxial layer thickness of \num{18}~and~\SI{25}{\um} were studied. In addition, the reverse substrate bias voltage was swept from 0~to~\SI{-6}{\V}.

\begin{table}[!htb]
\caption{Summary of the pixel geometries analysed in this study. The pixel pitch is \SI{28}{\um} and variants were produced on~\SI{18}{\um}- and~\SI{25}{\um}-thick epitaxial layers.}
\footnotesize
\centering
\begin{tabular}{l|c|c|c|c|c|c|c|c|c|c|c|c|c}
Matrix \# & 69  & 70  & 73  & 74  & 75$^\star$  & 76  & 77  & 79  & 80  & 84  & 85  & 89  & 90  \\
\hline
N-well \si{(\micro\metre)} & \multicolumn{2}{c|}{1.2} & \multicolumn{5}{c|}{2}   & \multicolumn{2}{c|}{3}   & \multicolumn{2}{c|}{4}   & \multicolumn{2}{c}{5}   \\
\hline
Spacing \si{(\micro\metre)} & 2   & 3   & 1   & 2   & 3   & 4   & 5   & 2   & 3   & 2   & 3   & 2   & 3 
\end{tabular}\\[1ex]
 \footnotesize $\star$: ALPIDE-like reference pixel for which also timing and different reverse substrate biases were studied.
\label{tab:geomtries}
\end{table}

\subsection{Circuitry and mode of operation}
\label{sec:circuitry}

\begin{figure}[thp]
	\centering
    \begin{subfigure}[t]{0.62\textwidth}
        \includegraphics[width=\textwidth]{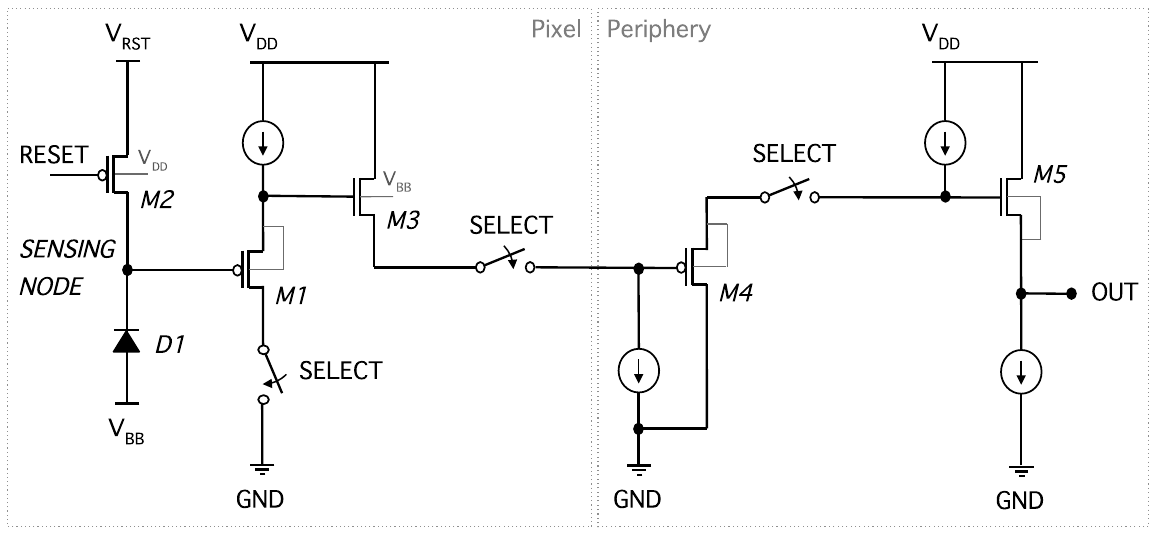}
        \caption{simplified schematic}
        \label{fig:schematic}
    \end{subfigure}
    \begin{subfigure}[t]{0.37\textwidth}
	    \includegraphics[width=\textwidth]{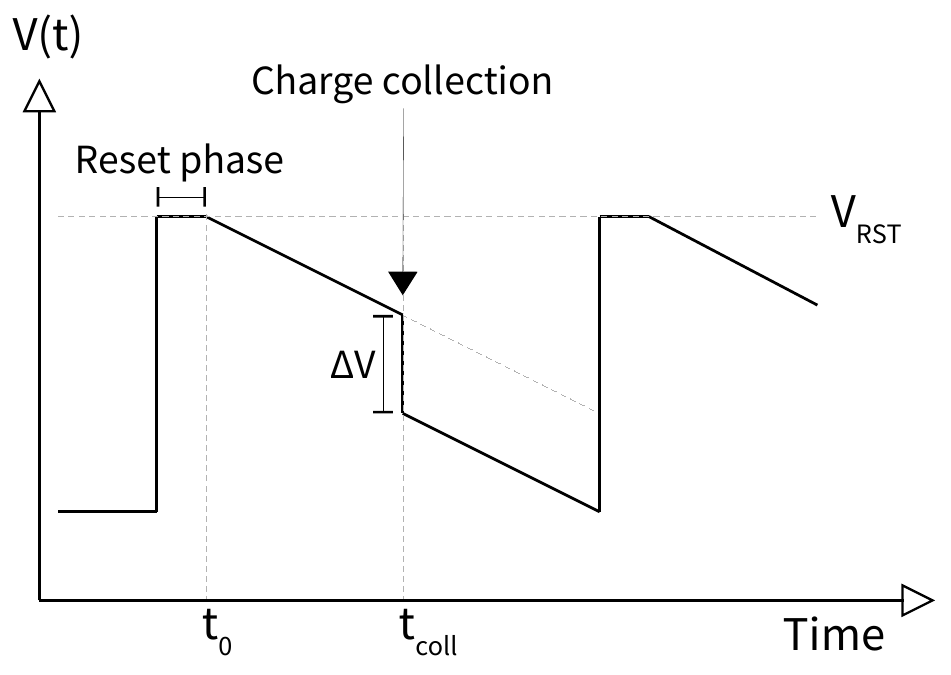}
	    \caption{input node potential vs time}
	    \label{fig:sequence}
	\end{subfigure}
	\caption{(a) Circuitry of the pixels discussed in this paper along with the matrix selection and output buffer circuitry. (b) Output voltage signal as a function of time. The input node potential is brought to $V_{RST}$ by closing~\texttt{M2} (reset phase) and is slowly decreased by the leakage (enlarged for illustrative purposes). A charge collection is observed as a potential drop $\Delta V$ on~\texttt{D1} which is brought to the output pin.}
\end{figure}

Figure~\ref{fig:schematic} shows the circuitry used to control and read out the ``active reset'' pixels studied in this paper. The chain of multiplexers and buffers in the periphery selects the matrix to be looked at. The pixel circuitry is instantiated \num{64}~times, to read out the central \num{64}~pixels of a \num{10x10}-pixel matrix simultaneously.
The source follower~\texttt{M1} acts as a buffer that isolates the sensing diode from the readout circuit, which is formed by \texttt{M3}--\texttt{M5}. The overall gain of this chain is slightly below unity, but uniform (within \SI{\approx1}{\percent}) across pixels. The multiplexing scheme eventually limits the bandwidth of the Investigator circuit to below \SI{100}{\MHz}.

The signal formation and acquisition is illustrated in Fig.~\ref{fig:sequence}, and works as follows:
\begin{enumerate}
    \item During the reset phase the PMOS~\texttt{M2} is activated and the pixel diode is charged to~$V_\text{RST}=\SI{0.8}{\V}$.
    \item The pixel potential decreases slowly due to the leakage of~\texttt{D1}.
    \item If a charge~$Q$ is collected the potential drops as~$\Delta V=Q/C$, with $C$~being the pixel capacitance.
    \item The potential continues to decrease due to leakage.
    \item The next reset phase is entered in a periodic fashion.
\end{enumerate}

The duration of the reset phase and the repetition frequency of the sequence are tuned such that the reset is long enough to fully restore the nominal potential and that the leakage does not change the potential significantly in absence of particle hits\footnote{The reset phase duration was set to \SI{3}{\us}, the acquisition window was opened \SI{9.2}{\us} after the reset phase ended and it lasted for \SI{15.7}{\us}. During the acquisition window, in absence of particle hits, the average observed voltage reduction was \SI{<1}{\mV}.}.

\section{Measurement set-up}
\label{sec:setup}

\subsection{Readout system}
\label{sec:invros}
The Investigator readout system~\cite{jacobus_thesis} features \num{64}~\SI{14}{-bit} ADCs with \SI{65}{\MHz} sampling rate, allowing the parallel readout of the central \num{8x8} pixels in a matrix. It is operated in free-running mode, where reset pulses are given periodically to the pixels. Between the resets the signal is continuously monitored for a sudden potential drop in any of the \num{64}~channels. When this trigger event\footnote{More technically, the acquisition trigger is given by difference between two consecutive ADC samples being larger than defined threshold; in this paper \SI{100}{ADC counts} or approximately \SI{12}{\mV}.} happens, the  acquisition window of 1024 samples between the two reset pulses is stored for all channels.

\begin{figure}[!ht]
	\includegraphics[width=\textwidth]{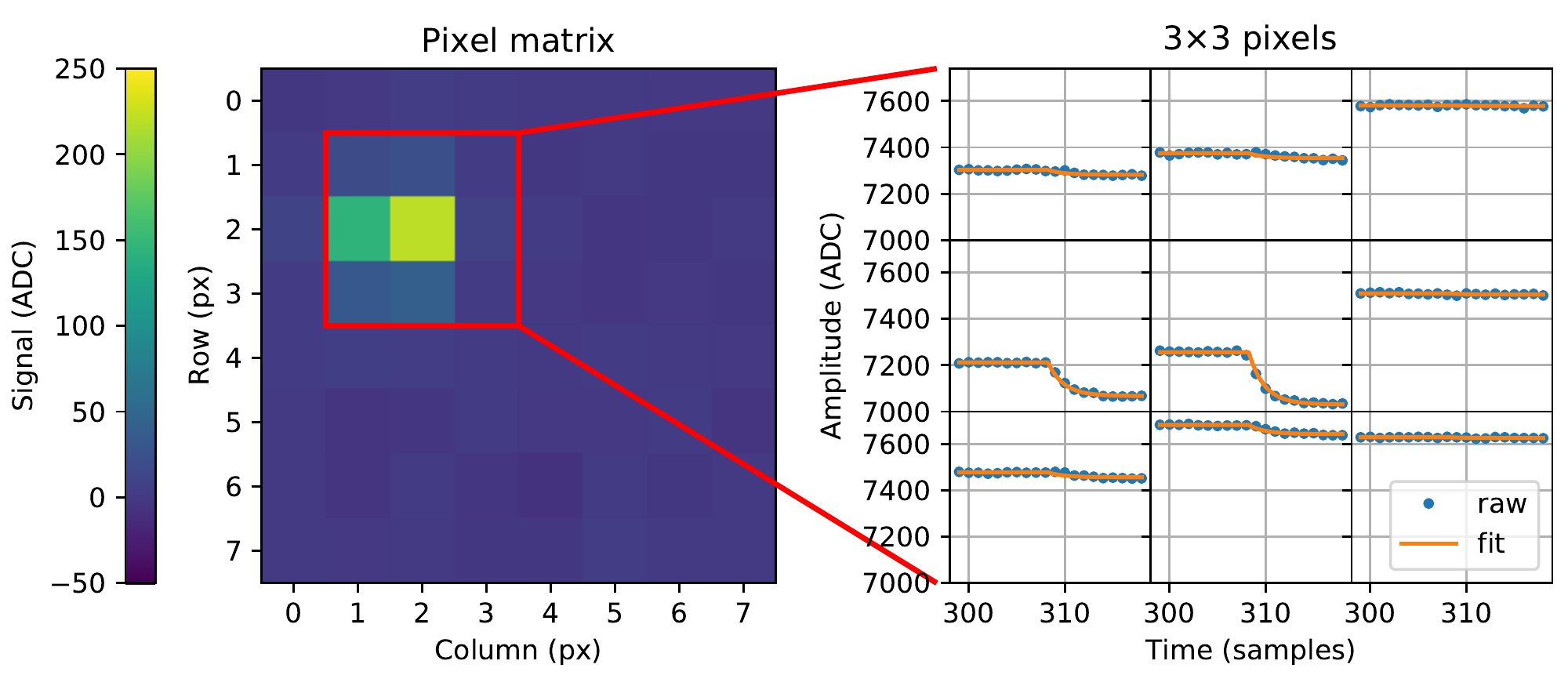}
    \caption{Example of a raw Investigator event showing the extracted signal for the full \num{8x8}-pixel matrix as well as a zoom of a \num{3x3}-pixel matrix (marked in red) with \num{\pm10} samples around the trigger time (here at sample~\num{309}). The geometry and bias parameters for this example are: \SI{25}{\um}-thick epitaxial layer, \SI{2}{\um}-wide diode n-well, \SI{3}{\um} spacing, \SI{-1}{\volt} reverse substrate bias. One ADC count corresponds to roughly \SI{0.12}{\mV} and one time bin to \SI{15}{\ns}. For a description of the fit see Sec.~\ref{sec:timing}.}
	\label{fig:rawevent}
\end{figure}

Figure~\ref{fig:rawevent} shows an example event. The signal of a pixel is defined as the voltage step observed at the moment of charge collection. It is extracted from the recorded wave forms as the difference of the averages of samples before and after the triggering voltage step. The fact that the charge collection can last for several sampling periods is taken care of by sampling sufficiently far away from the triggering sample (\SI{>30}{samples}).
In the event in Fig.~\ref{fig:rawevent}, the conversion presumably happened close to the boundary of two pixels and the generated charge is shared among them. The collection is rather slow, which can be attributed to a contribution from diffusion.

The following quantities are extracted from the recorded signals:
\begin{itemize}[label={--}]
\item\textbf{cluster:} the set of edge-adjacent pixels, which exceeded a given threshold.
\item\textbf{cluster size:} the number of pixels with signal above \SI{100}{\ele} contributing to a cluster\footnote{The value of \SI{100}{\ele} is chosen to allow for a direct comparison with other, both analogue and digital, MAPS.},
\item\textbf{seed pixel:} the pixel with the largest collected charge in a cluster, e.g.~pixel $(2,2)$ in Fig.~\ref{fig:rawevent},
\item\textbf{seed signal:} the charge collected by the seed pixel,
\item\textbf{matrix signal:} the total charge collected in a \num{3x3}-pixel matrix centred at the seed signal (see~Fig.~\ref{fig:rawevent}).,
\item\textbf{one-pixel cluster:} a cluster with only one pixel with signal above threshold, and all neighbour pixels with signal below \SI{1}{\mV} (two times the electronic noise),
\item\textbf{rise time constant:} time constant of the fit to an exponential of a pixel signal (see Fig.~\ref{fig:rawevent} and Sec.~\ref{sec:timing}).
\end{itemize}

\subsection{Fe-55 source}
\label{sec:fe55}
An \isotope[55]{Fe} source was used for the measurements carried out in this paper. It produces two characteristic soft X-rays 
at around~\SI{5.9}{\keV} ($K_\alpha$) and~\SI{6.5}{\keV} ($K_\beta$) to which MAPS are sensitive~\cite{radionuclides}. 

It is a convenient source for these studies: the deposited charge is comparable to that of minimum ionising particles traversing sensitive layers of \SIrange{20}{30}{\um}, and the absorption probability is adequate (absorption lengths of \SI{\approx30}{\um}~\cite{nist2}), while at the same time not introducing a too strong non-uniformity.
The main interaction process for these X-rays in silicon is photoelectric absorption~\cite{nist2}. An electron emitted by the photoelectric absorption at those energies, generates electron-hole pairs within~\SI{1}{\micro\metre} of its origin~\cite{xdb}, which can be treated here as point-like charge deposition.

The two X-ray energies are known to release on average \SI{1640}{\ele} and~\SI{1800}{\ele} in silicon, respectively, and can hence be used to calibrate the sensor signal in terms of charge. In this respect, the $K_\alpha$-peak at~\SI{1640}{\ele} is referred to as ``calibration peak''. A charge calibration is performed to be able to better compare different pixel geometries and bias settings in terms of charge sharing; note also that a threshold of~\SI{100}{\ele} is used to define the cluster size.

The observed signal shape of an \isotope[55]{Fe}~X-ray depends on the position where its absorption takes place with respect to the geometry of the detector. Three primary regions of interest can be identified:
\begin{itemize}
  \item\textbf{High electric field region}: The electron-hole pairs are created inside the high electric field region below a collection electrode. They are directly collected by drift, resulting in a one-pixel cluster.
  \item\textbf{Low or no electric field region in the epitaxial layer}: The charge transport mechanism is dominantly thermal diffusion until the carriers reach a high electric field region and are collected (or until they recombine). The deposited charge can be shared between several pixels due to the statistical nature of the diffusion process.
  \item\textbf{Substrate:} Electrons diffuse thermally in the substrate. Their lifetime before recombination is small, such that only a fraction reaches the epitaxial layer. The part reaching the epitaxial layer is collected as above.
\end{itemize}
In the first two cases, the entire charge deposited by the photo electron is expected to be collected, given the relatively long carrier lifetime with respect to the collection time. In the third case the amount of collected charge depends on the depth of the photoelectric conversion and the carrier lifetime in the substrate.

The carrier lifetime varies by orders of magnitude from low-doping (epitaxial layer) to high-doping regions (substrate), i.e.~changes from above~\SI{10}{\us} to below~\SI{20}{\ns}, respectively\footnote{Values estimated using $\tau (N_A) = \left( 1+\frac{N_A}{N_{\mathrm{ref}}} \right)^{-1} \tau_{\mathrm{max}}$ \cite{fossumlee}, with $\tau_{\mathrm{max}}=\SI{e-5}{\s}$, $N_{\mathrm{ref}}=\SI{e16}{\per\cubic\cm}$, and $N_A$ equal to \SI{9e11}{\per\cubic\cm} and \SI{5e18}{\per\cubic\cm} in the epitaxial layer and the substrate, respectively.}. Furthermore, the diffusion coefficient in the substrate is an order of magnitude lower than in the epitaxial layer~\cite{mobility}.
Therefore, only a limited amount of electrons generated in the substrate is collected. In particular, only the photons converted in the first \SI{5}{\um} of the substrate are measurable~\cite{simulation}.

\begin{figure}[thp]
\includegraphics[width=\linewidth]{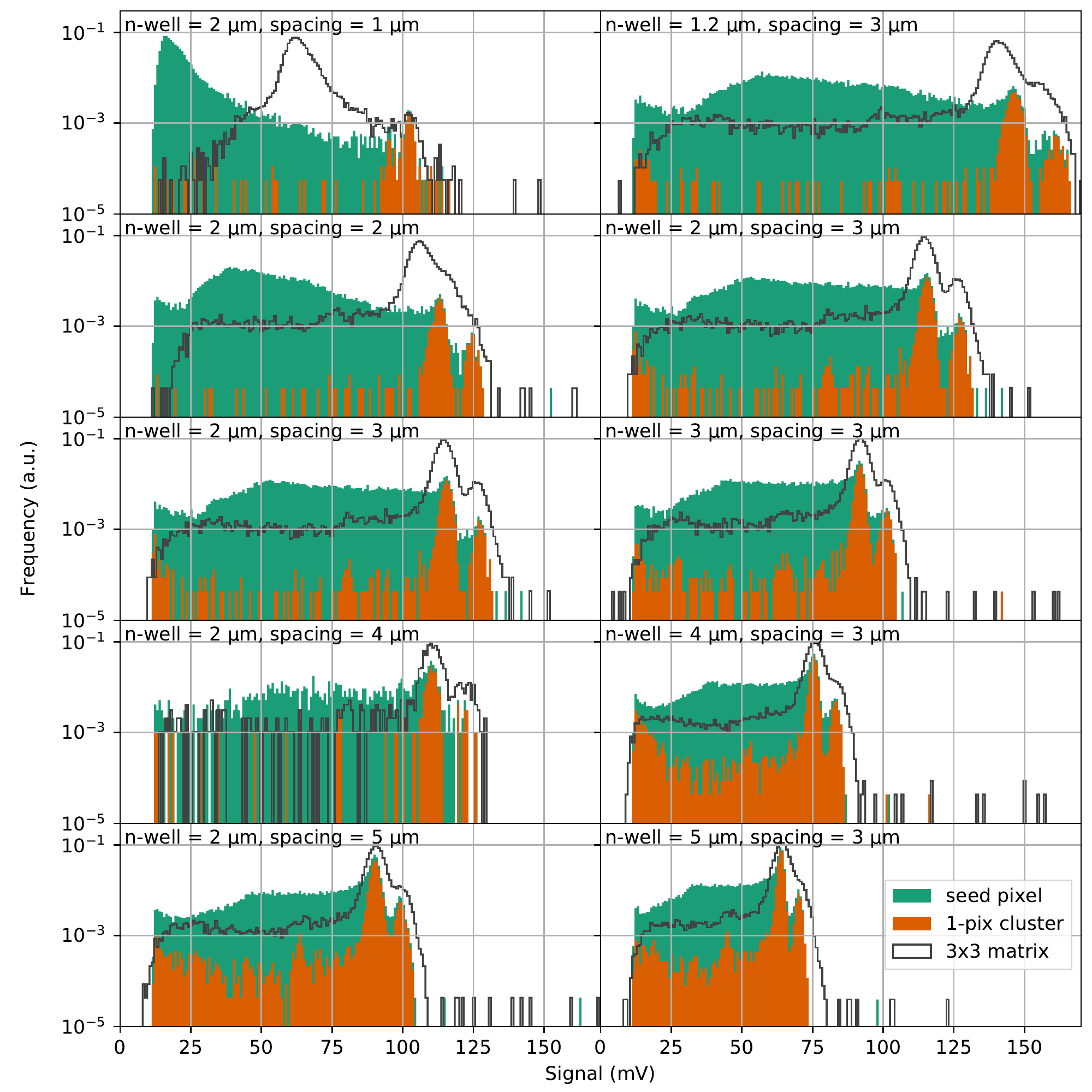}
\caption{Example Fe-55 spectra for nine different diode geometries on a \SI{25}{\um}-thick epitaxial layer and reverse substrate bias of~\SI{-6}{\V}. The left column shows the effect of changing the spacing, the right the effect of changing the diode n-well size.}
\label{fig:spectra}
\end{figure}

Figure~\ref{fig:spectra} shows a selection of recorded signal spectra. Here, both the \num{3x3}-pixel matrix signal and the seed pixel signal  are shown. The latter is further detailed into cases where the cluster size is one. The double-peak structure from the $K_\alpha$ and $K_\beta$ \isotope[55]{Fe} X-rays is immediately visible, in particular in the matrix and best in the one-pixel spectra.
The seed pixel signal spectra show an increased frequency at around one third to one half of the $K_\alpha$-peak, which is attributed to the events where charge is shared between pixels. The matrix spectra contain a long tail originating from X-ray conversions in the substrate, where the generated electrons are only partially captured.

\section{Results}
\label{sec:results}
The measurements are split in two categories, first the asymptotic time behaviour of the charge collection is studied, then, the time response is studied for a selected pixel geometry.

\subsection{Asymptotic charge collection}
\label{sec:asymptotic}

\begin{figure}[!htb]
\includegraphics[width=\linewidth]{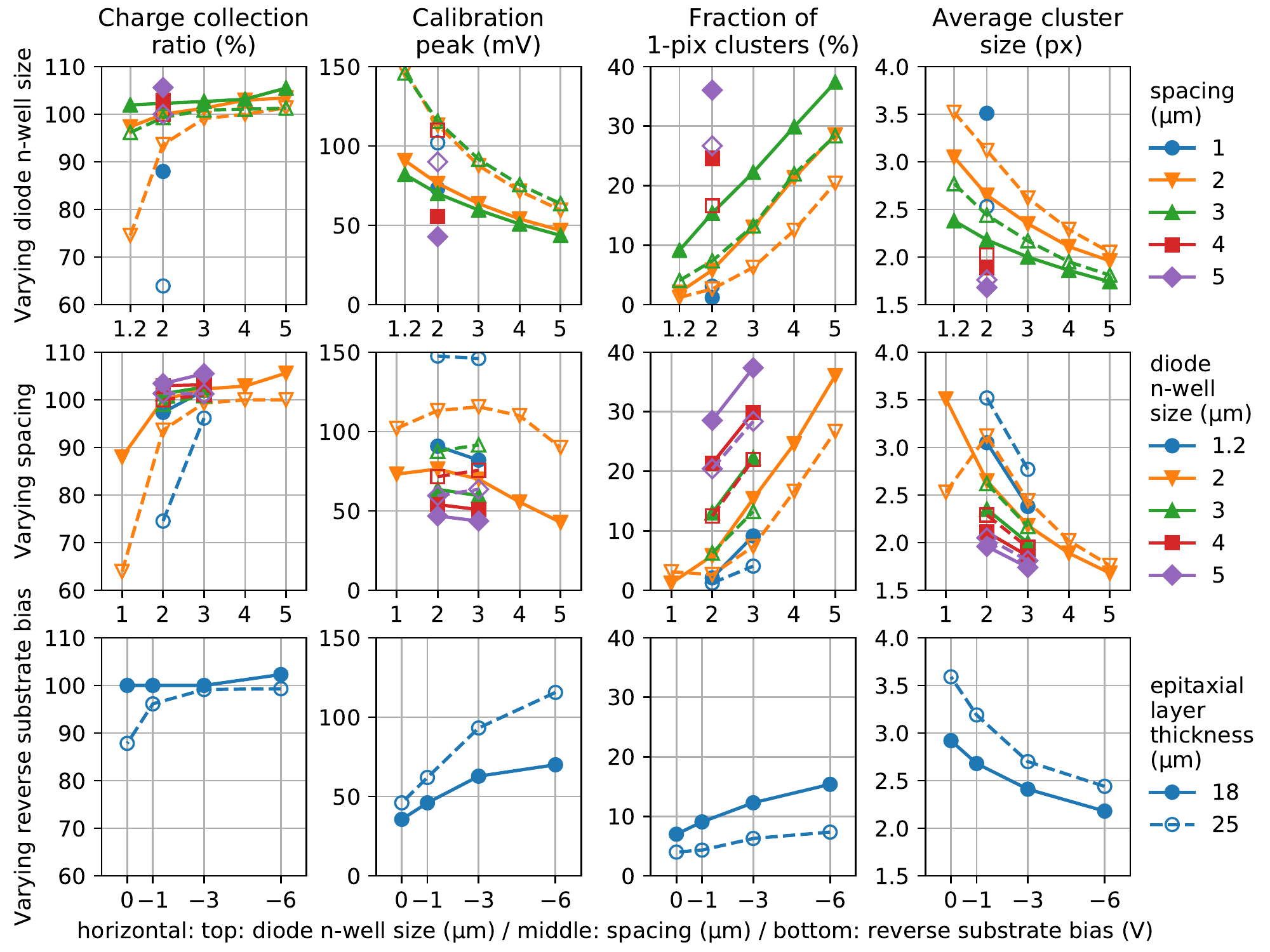}
\caption{Extracted asymptotic charge collection parameters. The two top rows show the dependence on geometric parameters for a fixed reverse substrate bias of~\SI{-6}{\V}, the bottom row shows the influence of the reverse substrate bias for a fixed geometry with diode n-well size of~\SI{2}{\um} and spacing of~\SI{3}{\um}. All figures contain data from epitaxial layer heights of~\SI{18}{\um} (solid lines, closed symbols) and \SI{25}{\um} (dashed lines, open symbols). See text for the definition of the parameters.}
\label{fig:static}
\end{figure}

The following quantities are extracted from the \isotope[55]{Fe}  measurements and are summarised in~Fig.~\ref{fig:static}:
\begin{itemize}
    \item\textbf{Charge collection ratio:} The ratio of the most-probable value of the \num{3x3}-pixel matrix signal distribution to the most-probable value of the one-pixel cluster signal distribution. It is an indicator of the charge collection efficiency\footnote{The intrinsic assumption here is that in a one-pixel cluster the entire deposited charge is collected by one pixel. However, as a small amount of charge can be still be collected by other pixels (visible in slightly skewed one-pixel cluster distributions in Fig.~\ref{fig:spectra}), thus resulting in a ratio above \SI{100}{\percent}.}.
    \item\textbf{Calibration peak:} The amplitude of the calibration peak (expressed as voltage) seen in the seed signal distribution (the~\SI{1640}{\ele} from the \SI{5.9}{\keV}~X-ray). It is an indication\footnote{It is an indication only, as the voltage gain of the circuit is not known precisely enough.} of the pixel capacitance~($C=Q/\Delta V$).
    \item\textbf{Fraction of one-pixel clusters:} The relative number of clusters of size one with respect to the total number of clusters. This quantity indicates the relative spatial extension of high electric field volume of a pixel (see~Sec.~\ref{sec:fe55}).
    \item\textbf{Average cluster size:} A higher average cluster size is an indication of a higher charge sharing between pixels.
\end{itemize}

Several trends can be observed, most of which can qualitatively be described by the change of the depletion volume i.e.~the electric field and associated changes of capacitance and charge collection mechanism:
\begin{enumerate}
    \item It can be clearly observed that the extreme choices of smallest diode n-wells or smallest spacing yield bad results in the sense that charges are being lost.
    \item Increasing the diode n-well size or the spacing lead to different trade-offs between signal amplitude and charge sharing. Here it is worth mentioning that \emph{some} charge sharing is typically (and certainly for the ALICE application) desirable as it can increase the spatial resolution of the sensor. However, charge sharing also divides the charge over several pixels thus putting higher requirements on the ability to resolve smaller signal and leading to a lower radiation hardness.
    \item A thicker epitaxial layer leads to more charge sharing and needs a more careful tuning of the geometry. For tracking charged particles, however, it has the advantage that the induced signal is larger (scales with path length).
    \item Reverse substrate bias has a very large influence on the performance of the sensor, generally being beneficial.
\end{enumerate}
In these observations, a larger spacing leads to two competing mechanisms onto the capacitance of the input node: the well capacitance decreases due to a larger depletion, but the capacitance of the metal line connection of the diode n-well to the first input transistor (residing in the deep p-well) increases as it becomes longer. 

The trade-off made for the ALPIDE chip for ALICE Inner Tracker System is a \SI{2}{\um}-wide well with a \SI{3}{\um}~spacing on a \SI{25}{\um} epitaxial layer. The bias voltage is adjustable~\cite{ALPIDE-proceedings-1, ALPIDE-proceedings-2,ALPIDE-proceedings-3}.

\subsection{Time-resolved charge collection}
\label{sec:timing}
The Investigator chip and its readout system allow to resolve the time response of a pixel down to values around \SI{15}{\ns}, mostly determined by the bandwidths of the chip and the readout system as well as the sampling rate. 
In the absence of an external trigger in the measurement of X-rays, the charge collection time is defined as the time constant $\tau$ extracted from a fit of an exponential step-response to the signal:
\begin{align}
    V(t)=
    \begin{cases}
    V_0 & t<t_0\\
    V_0-\Delta V\cdot[1-\exp(-(t-t_0)/\tau)] & t\ge t_0
    \end{cases}\quad.
\end{align}
Such a fit is for example shown in the right-hand panel of Fig.~\ref{fig:rawevent}. It should be noted that this fitting procedure allows to obtain $t_0$ and $\tau$ with sub-sample resolution. Also note that the time $T_{\SI{90}{\percent}}$ to collect \SI{90}{\percent} of total collected charge would be equivalent to $\approx2.3\,\tau$ in this model.

\begin{figure}[!htb]
\includegraphics[width=\linewidth]{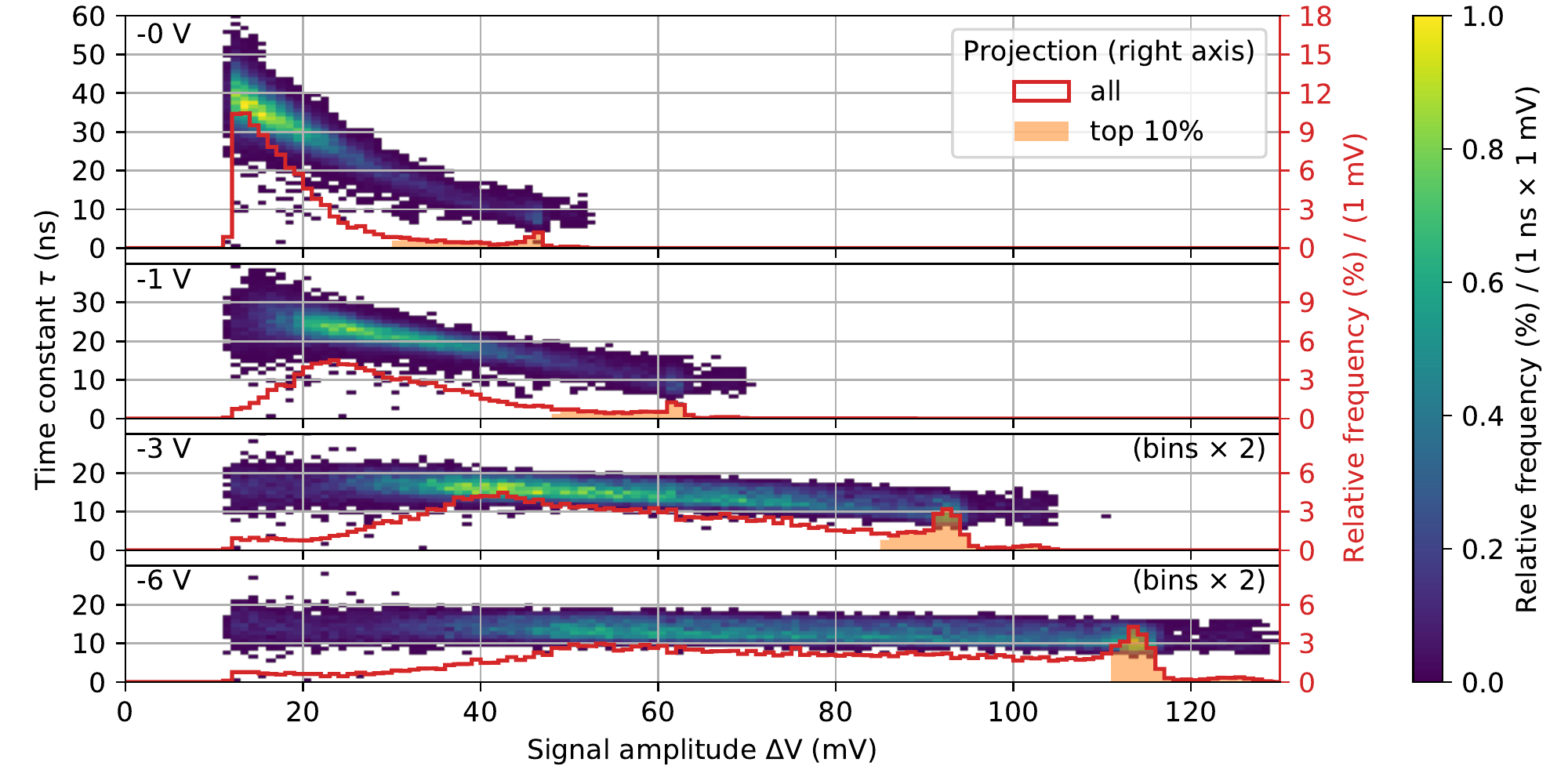}
\caption{Time response of seed signal versus its amplitude for different values of reverse substrate bias for a pixel with diode n-well size of~\SI{2}{\um} and spacing of~\SI{3}{\um} on a \SI{25}{\um}-thick epitaxial layer. The right-hand vertical axis shows the projection onto signal amplitude, showing the seed signal spectrum. The highlighted part corresponds to the \SI{10}{\percent}-quantile of signals with the highest amplitude.} 
\label{fig:taus}
\end{figure}

Figure~\ref{fig:taus} shows the extracted time constants of seed pixel signals for different values of reverse substrate bias and signal amplitude. It can be observed that there are no measured time constants below \SI{\approx 8}{\ns}, due to the bandwidth-induced limit of the system, but many signals are hitting this limit.  This result, the time constant equivalent to the system bandwidth, can be regarded as the main result already -- it shows that the dominant charge collection process is drift rather than diffusion. In particular the calibration peaks are collected very fast, confirming the previous assumption that the calibration peak originates mostly from the collection by drift.

\begin{figure}[!htb]
\includegraphics[width=\linewidth]{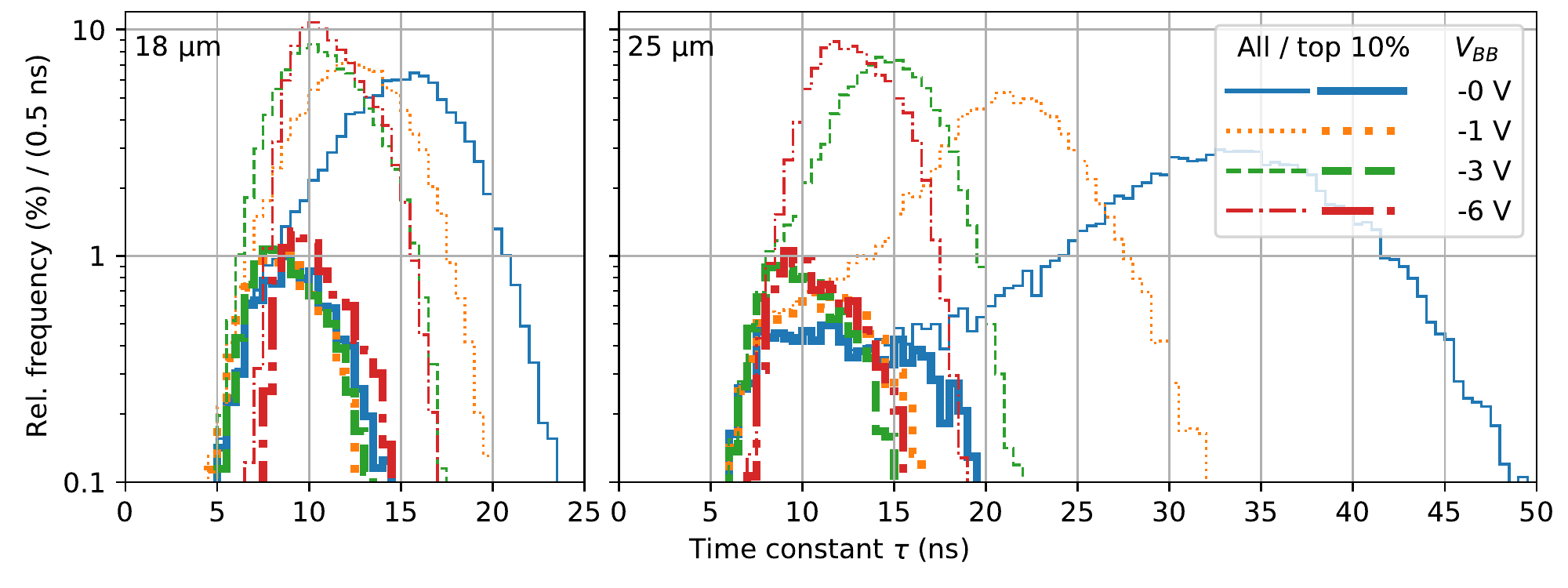}
\caption{Distribution of signal time constants for different reverse substrate bias voltages and epitaxial layer thicknesses. The thick lines correspond to the \SI{10}{\percent}-quantile of signals with the highest amplitude, which can be largely attributed to collection within the high electric field volume (cf.~Fig.~\ref{fig:taus}). The studied pixel has a diode n-well size of~\SI{2}{\um} and a spacing of~\SI{3}{\um}.}
\label{fig:timing}
\end{figure}

Figure~\ref{fig:timing} compares the time responses for different reverse substrate bias voltages and epitaxial layer thicknesses, also indicating the time response of the \SI{10}{\percent}-quantile of signals with the highest amplitude, corresponding to the area where the seed signal is found. The following observations can be made:
\begin{enumerate}
    \item The time constant of the calibration peak is at the minimum measurable value given by the system resolution (i.e.~\SI{<10}{\ns}) for all cases.
    \item A thicker epitaxial layer needs more reverse substrate bias to reach high collection speeds throughout the full volume. The part collected in the high electric field volume (calibration peak) shows comparable speeds (below system resolution).
    \item The contributions to the signal attributed formerly to the low electric field part of the sensor become faster with more bias voltage, eventually reaching the system resolution over their full range at $V_{BB}=\SI{-6}{\V}$.
\end{enumerate}

\section{Summary}
\label{sec:summary}
The parametric study of different pixel designs shows the influence of geometric parameters, of the choice of epitaxial layer and of the bias conditions. While the observed trends can be understood qualitatively by simple arguments, their quantitative importance is more involved. The reported measurements can, hence, serve as reference for choosing design parameters in future developments and as benchmark of more detailed simulations.

In particular, it is worth emphasising the influence of adding an undoped gap (spacing) between the diode n-well and the surrounding p-well. It turns out to be the key parameter to largely improve the performance at a relatively small penalties in terms of capacitance increase and area requirement.

The presented results show that fast charge collections are obtainable in these kind of sensors.
This naturally raises an interest of applications requiring $O(\si{\ns})$-time resolutions or, given that the dominant charge collection mechanism is drift, radiation hardness. Eventually, in order to time-resolve the signal generation and collection of these type of sensors, a new chip and readout system with larger bandwidth would be required in the future.

%


\section*{References}

\bibliography{main.bib}

\end{document}